\documentclass[12pt]{article}

\begin{document}

\title{Field-theoretic methods for systems of particles with exotic
exclusion statistics}

\author{
A.~S.~Stepanenko and J.~M.~F.~Gunn
\\
{\small\it School of Physics and Space Research, University of Birmingham,}
\\
{\small\it Birmingham B15 2TT, United Kingdom.}
}

\date{ }

\maketitle

\begin{abstract}
We calculate the partition function of a gas of particles obeying
Haldane exclusion statistics, using a definition of a Hilbert space
having a `fractional dimension' and constructing appropriate coherent
states.  The fractional dimension is expressed though the form of the
identity operator in the Hilbert space. 

We find that there many possible generalisations of the Pauli exclusion
principle, with particular choices of the scalar product leading to
consistency either with Haldane's original definition of the effective
dimensionality of the Hilbert space or with the combinatorial procedure
invoked by Haldane and Wu. We explicitly demonstrate that at low
particle densities these definitions are equivalent.

\end{abstract}

\section{Introduction}

Haldane introduced in~\cite{Hald} a generalized exclusion principle defining a
quantity $d(N)$, the Haldane dimension, which is the dimension of the
one-particle Hilbert space associated with the $N$-th particle, keeping the
coordinates of the other $N-1$ particles fixed.  The statistical parameter, $g$,
of a particle (`$g$-on') is defined by (where we add $m$ particles)
\begin{equation}
g = - \frac{d(N+m) - d(N)}{m}
\end{equation}
and the conditions of homogeneity on $N$ and $m$ are imposed. The system
is assumed to be confined to a finite region where the number $K$ of
independent single-particle states is finite and fixed. Here the usual
Bose and Fermi ideal gases have $g=0$ for Bose case (i.e. $d(N)$ does not
depend on $N$) and $g=1$ for Fermi case -- that is the dimension is
reduced by one for each added fermion, which is the usual Pauli
principle.

Haldane also introduced a combinatorial expression (which we will term the
Haldane-Wu state-counting procedure~\cite{Hald,Wu}) for the number of ways,
$W$, to place $N$ $g$-ons into $K$ single--particle states. Then
\begin{equation}
W = \frac{[d(N)+N-1]!}{[d(N)-1]! N!}  \qquad d(N)=K-g(N-1) ,
\label{W}
\end{equation}
which was subsequently used by many authors \cite{MS1,WB,W2,Ha,Is,Raj} to
describe thermodynamical properties of $g$-ons. In particular Bernard and
Wu \cite{WB} and Murthy and Shankar \cite{MS2} showed that the behavior
of the excitations in the Calogero--Sutherland model is consistent with
Eqn.(\ref{W}) \cite{Wu} for $g$-ons, with fractional $g$, in general.

In Ref~\cite{Poly1} the microscopic origin of the Haldane-Wu
state-counting procedure was examined. The notion of
statistics was considered in a probabilistic spirit.  The author assumed
that a single level may be occupied by any number of particles, and each
occupancy is associated with an a priori probability. These probabilities
are determined by  enforcing consistency with the Haldane-Wu
state-counting procedure and not with Haldane's definition of exclusion
statistics. There was no construction of a Hilbert space and the a priori
probabilities may be negative. This approach has been further elaborated
in a number of papers~\cite{Is2,Chat,MS3}.

Another probabilistic approach has been developed in \cite{IG,IIG1,IIG2}.  It
was pointed out that there is a distinction between Haldane's dimension and
Haldane-Wu state counting procedures.  A `fractional' Hilbert space (associated
with the non-integer nature of $d(N)$) and the corresponding
creation-annihilation operators were constructed and a set of probabilities
which give Haldane's dimension was obtained.

The paper is organized as follows.  In the next section we introduce the notion
of a fractional Hilbert space and creation-annihilation operators associated
with it.  In section~3 we obtain a generalised resolution of unity in terms of
coherent states.  In section~4 the definition of Haldane's dimension is
considered in detail.  We calculate the partition function and the
state-counting expression.  In section~5 we consider the Haldane-Wu
state-counting procedure and make a comparison with the definition of Haldane's
dimension.

\section{Hilbert space and creation-annihilation operators}

In this section we recall the main ideas introduced in Ref~\cite{IIG1}.

The definition of a fractional dimension Hilbert space is connected with 
state-counting, which we need to calculate the entropy and other thermodynamical 
quantities of $g$-particles.

The main idea is to consider the process of inserting the $N$-th particle into 
the system as a probabilistic process (in spirit of Gibbs), i.e. we assume 
that the probability of such insertion plays the role of Haldane's measure 
of the probability to add the $N$-th particle to the system. Let us 
illustrate the idea for the case of a single degree of freedom, $g=1/p$ , 
and provide an interpretation of $d(N)$ for that case.

Firstly, we have the vacuum state to which we add the first particle.  We assume
that the nature of the statistics reveals itself at the level of two particles,
so $d(1)=1$.  Now let us assume that the process of insertion of the second
particle is a probabilistic one with the probability $(1-g)$ of success.  We
interpret this as fractional dimension, $d(2)$, of the subspace (corresponding
to double occupation) and $d(2)=1-g$. The conditional probability to add a third 
particle (with two assumed present) is $1-2g$. Hence the probability of success 
in adding three particles is $1\times(1-g)(1-2g)$. This leads us to the 
probability of adding $n$ particles is:
$$
\alpha_n = [1-g][1-2g]\cdots [1-(n-1)g]
$$
We see that the probability to find $N>p$ particles in the system is equal to 
zero.

Drawing parallels with dimensional regularization we can formulate a geometrical 
definition of the fractional dimension. In that case the trace of the
identity matrix is identified with the value of (non-integer) dimension, $d(N)$. 
In the calculation of thermodynamical quantities such as the partition 
function or the mean value of an arbitrary operator $\hat O$ we must 
compute the following traces:
\begin{equation}
Z = {\rm Tr}\left[{\rm Id}\cdot{\rm e}^{-\beta H}\right]\ ,\qquad
\langle\hat O\rangle
 = \frac{1}{Z}{\rm Tr}\left[{\rm Id}\cdot{\rm e}^{-\beta H}\hat O\right]
\end{equation}
where the Hamiltonian $H$, e.g. for an ideal gas, is 
\begin{equation}
H = \sum_{i=1}^K \epsilon_i n_i
\end{equation}
and the ``unit operator'', Id, which completely defines the exclusion 
statistics of the particles is defined by
\begin{equation}
{\rm Id} = \sum_{n_1,\dots,n_K=0}^\infty \alpha_{n_1,\dots,n_K}
|n_1,\dots,n_K\rangle\langle n_K,\dots,n_1|
\end{equation}
where $\alpha_{n_1,\dots,n_K}$ is the probability to find the state 
$|n_1,\dots,n_K\rangle$. Then the full dimension of the $N$-particle 
subspace is given by the formula
\begin{equation}
W(N)={\rm Tr}\left(\left.{\rm Id}\right|_{\sum_{i=1}^K n_i = N}\right)
= \sum_{n_1+\dots+n_K = N}\alpha_{n_1,\dots,n_K}
\label{W(N)}
\end{equation}
An analog of Haldane's dimension, $d(N)$, for the $N$-particle subspace with 
an arbitrary fixed $(N-1)$-particle substate is described by the relation
\begin{equation}
d(N) = \sum_{\ell=1}^N\frac{\alpha_{n_1,\dots,n_{\ell}+1,\dots,n_K}}
{\alpha_{n_1,\dots,n_K}|_{\sum n_i = N-1}}
\label{d(N)}
\end{equation}
The above procedure is completely general, a concrete choice of the 
probabilities $\alpha_{n_1,\dots,n_K}$ is not required.

On the basis of the following two assumptions:
\begin{enumerate}
\item the definition of the $N$-th particle dimension $d(N)$ actually 
yields Haldane's conjecture $d(N) = K-g(N-1)$;

\item the Hilbert space of the system with $K$ degrees of freedom is 
factorized into a product of Hilbert spaces corresponding to each degree of 
freedom. This means
$$
{\rm Id} = {\rm Id}_1\otimes{\rm Id}_2\otimes\dots\otimes{\rm Id}_K
$$

\end{enumerate}
it was shown, in Ref~\cite{IIG1}, that there is a single self-consistent way to 
define $\alpha_{n_1,\dots,n_K}$:
\begin{equation}
\alpha_{n_1,\dots,n_K} = \prod_{i=1}^K [1-g][1-2g]\dots [1-(n_i-1)g]
\end{equation}
where $g=1/p$, $p$ integer. In this case the statistical parameter $g$ can take 
values between $0$ and $1$.

If we weaken the second condition and allow matrix Id to be a direct product 
of Id's which correspond to some elementary `exclusion cell' (block) with 
dimension $q>1$ but keeping the first condition, we obtain the following 
set of probabilities:
\begin{equation}
\alpha(\{n_{ij}\}_{i,j=1}^{K,q})
 = \prod_{i=1}^K \left[1-\frac{1}{p}\right]\left[1-\frac{2}{p}\right]\dots 
 \left[1-\frac{1}{p}\left(\sum_{j=1}^q n_{ij}-1\right)\right]
\label{alpha1}
\end{equation}
with statistical parameter $g=q/p$, $p$ integer and $K'=qK$ the full number of 
single particle states~\cite{IG}. Note that the statistical parameter can take 
values greater than 1.

Next we can weaken in addition the first condition and allow $d(N)$ to be a
non-linear function of $N=\sum_{i=1}^K n_i$. Then we find a large variety of 
probabilities, among them there is a set of probabilities corresponding 
to the Haldane-Wu state-counting procedure.

To allow interactions between exclusons, `hopping' or interaction with some
random potential we should develop a second-quantized formalism.  Representation
for these operators can be found from the following conditions:
\begin{equation}
a_i^\dagger|n_1\dots n_i\dots n_K\rangle
 = \beta_{n_1\dots n_K} |n_1\dots n_i+1\dots n_K\rangle
\end{equation}
\begin{equation}
(a_i^\dagger)^\dagger = a_i = {\rm Id}^{-1} (a^\dagger)^* {\rm Id}
\end{equation}
\begin{equation}
N_i |n_1\dots n_K\rangle = n_i |n_1\dots n_K\rangle
\end{equation}

The most interesting case is a system consisting from one exclusion cell, 
that is $g=K/p$, $K$ the full number of single particle states. In this 
case coefficients $\beta$ depend only on $n_i$ and $n=\sum_k n_k$:
\begin{equation}
\beta_{n_1\dots n_K} = \beta_{n_i, n}
 = \sqrt{(n_i+1)\frac{\alpha_n}{\alpha_{n+1}}}
\end{equation}
A remarkable result is that for the hopping term ($a_i^\dagger a_j$) 
the dependence on $\alpha$ disappears and it can be represented as
\begin{equation}
a_i^\dagger a_j = b_i^\dagger b_j P(p)
\end{equation}
where $b^\dagger,b$ are the bosonic operators and $P(p)$ the projector 
on a subspace with the number of particles less than or equal to $p$.

\section{Coherent states}

To illustrate the idea consider the simplest case $K=1, g=1/p$ and exchange 
statistics between $g$-particles to be bosonic one. Let us confine ourself for a 
moment by considering Hamiltonians depending on the number of particles only. 
Then states $|n\rangle$ can be chosen to be bosonic ones:
$$
|n\rangle = \frac{1}{\sqrt{n!}} (a^\dagger)^{n} |0\rangle\ ,\qquad
n = a^\dagger a
$$
$ a_i^\dagger, a_i$ are bosonic operators.

There is a well-known expressions for a trace and a expansion of the unit in 
terms of the bosonic coherent states:
\begin{equation}
{\rm Tr}[\hat O] = \frac{1}{\pi}\int {\rm d}\bar z{\rm d} z\ {\rm e}^{-\bar z 
z}\langle\bar z|\hat 
O|z\rangle
\end{equation}
\begin{equation}
\mbox{I} = \frac{1}{\pi}\int {\rm d}\bar z{\rm d} z\ {\rm e}^{-\bar z 
z}|z\rangle
\langle\bar z| = \sum_n |n\rangle \langle n|
\end{equation}
\begin{equation}
|z\rangle = {\rm e}^{a^\dagger z}|0\rangle\ ,\quad
\langle\bar z| = \langle 0|{\rm e}^{a\bar z}
\end{equation}
If  we express the matrix Id in terms of the bosonic coherent states then 
we can directly apply the bosonic technique to the system of exclusons. 
Having a look on the usual resolution of the unit we can conclude that if we 
find a function $F$ such that
\begin{equation}
\frac{1}{\pi}\int {\rm d}\bar z{\rm d} z\ F(\bar z z)|z\rangle
\langle\bar z| = \sum_n \alpha_n|n\rangle \langle n| = {\rm Id}
\end{equation}
we solve the problem. Rewriting the probabilities in the form
$$
\alpha_n=\frac{p!}{p^n(p-n)!}
$$
it can be shown that the following relation for the matrix Id holds
\begin{equation}
{\rm Id} = \int_C {\rm d} t\ F_p(t)\ \int{\rm d}\bar z{\rm d} z\ 
{\rm e}^{-\bar z z}  |z t^{1/2}\rangle 
\langle\bar z t^{1/2}|
\end{equation}
\begin{equation}
F_p(t) = \frac{1}{2\pi i}p!{\rm e}^{pt} t^{-p-1} p^{-p}
\label{F}
\end{equation}
where the contour $C$ runs around the origin in the complex plane in the 
counter clockwise direction. Noting that
\begin{equation}
\int_C {\rm d} t \ F_p(t) t^k = \frac{p!}{p^k(p-k)!}
\label{Frule}
\end{equation}
we see that the partition function takes the form:
\begin{equation}
Z_{1/p} = \int_C{\rm d} t\ F_p(t)(1-t{\rm e}^{-\beta\epsilon})^{-1}
 = \sum_{k=0}^p \frac{p!}{p^k(p-k)!} {\rm e}^{-k\beta\epsilon}
\end{equation}
For fermions ($p=1$):
$$
Z_{\rm f} = Z_1 = 1 + {\rm e}^{-\beta\epsilon}
$$
To investigate the bosonic limit the following representation for the 
partition function is useful:
\begin{equation}
Z_{1/p} = \int_0^\infty{\rm d} t\ {\rm e}^{-t} 
\left[1 + \frac{t{\rm e}^{-\beta\epsilon}}{p}\right]^p
\end{equation}
when $p\to\infty$ (bosonic limit) we have
$$
Z_{1/p} \mathop{\rightarrow}_{p\to\infty} Z_\infty
 = \int_0^\infty{\rm d} t\ {\rm e}^{-t+t{\rm e}^{-\beta\epsilon}} 
 = \frac{1}{1-{\rm e}^{-\beta\epsilon}} = Z_{\mbox{\tiny b}}
$$

\section{Haldane's dimension procedure}

In this section we consider in detail the set of probabilities 
(\ref{alpha1}) with $K=1$, i.~e. one exclusion cell ($q$ is the number of 
states):
\begin{equation}
\alpha(\{n_{j}\}_{j=1}^{q})
 = \prod_{j=1}^{N-1}\left[1-\frac{j}{p}\right]
 = \frac{1}{p^N}\frac{p!}{(p-N)!}\ ,\qquad
N = \sum_{j=1}^q n_j 
\label{alpha2}
\end{equation}
From (\ref{alpha2}) and (\ref{d(N)}) we have obviously
$$
d(N) = q\left[1 - \frac{N-1}{p}\right] = q - g(N-1)\ ,\qquad
g = \frac{q}{p}
$$
The matrix Id in this case has the form
\begin{equation}
{\rm Id}
 = \int\limits_C\!{\rm d}t\ F_p(t)
 \frac{1}{\pi^q}\int\limits\!\prod_{j=1}^q\left[{\rm d}\bar{z}_j{\rm d}z_j\ 
 {\rm e}^{-\bar{z}_j z_j} \right]
|\{z_j\sqrt{t}\}_{j=1}^q\rangle\langle \{\bar{z}\sqrt{t}\}_{j=1}^q|
\label{Id2}
\end{equation}
with the same function $F_p(t)$ defined in (\ref{F}) and the usual bosonic 
coherent states.

To express trace of an operator in terms of bosonic coherent states we use
the following relation
\begin{equation}
{\rm Tr}[{\rm Id}\cdot \hat{O}]
 = \frac{1}{\pi^{q}} 
 \int\limits\! \prod\limits_{j=1}^q {\rm d}\bar{w}_j{\rm d}w_j\ 
 {\rm e}^{-\sum_j\bar{w}_j w_j} 
 \langle \{\bar{w}_j\}_{j=1}^q| {\rm Id}\cdot \hat O |\{w\}_{j=1}^q\rangle 
\label{tr}
\end{equation}
The last relation allows us to calculate a partition function of $g$-ons 
with Hamiltonian
\begin{equation}
\hat H = \epsilon \hat N\ ,\qquad
\hat N = \sum_{i=1\cdots q} n_{i} = \sum_{i=1\cdots q} a_{i}^\dagger a_i  
\label{Ham}
\end{equation}
with $a^\dagger_i,a_i$ being bosonic creation-annihilation operators. From 
(\ref{tr},\ref{Id2}) we have
$$
Z_{q/p} = 
\int\limits_C\!{\rm d}t\ F_p(t)
\frac{1}{\pi^{2q}} \int\!{\rm D}\bar{w}{\rm D}w
{\rm D}\bar{z}{\rm D}z\ {\rm e}^{-\bar{w}w-\bar{z}z}\
\langle \bar{w}|z\sqrt{t}\rangle
\langle \bar{z}\sqrt{t}| {\rm e}^{-\beta(\hat{H}-\mu\hat{N})} |w\rangle
$$
where 
$$
{\rm D}\bar{w}{\rm D}w \equiv 
\prod\limits_{j=1}^q{\rm d}\bar{w}_j{\rm d}w_j
$$
and summations in the exponential are implied.

Using the following relation
\begin{equation}
\exp[c a^\dagger a]
 = {\rm N}\left[ \exp\Bigl( ({\rm e}^{c}-1) a^\dagger a \Bigr)\right]
\label{norm}
\end{equation}
(here N stands for normal form of an operator expression) and the following 
properties of the coherent states:
\begin{equation}
\langle \bar{w}|{\rm N}Q(a^\dagger,a)|z\rangle
 = Q(\bar{w},z) \langle \bar{w}|z\rangle\ ,\qquad
\langle \bar{w}|z\rangle = \exp(\bar{w} z)
\label{prop}
\end{equation}
we obtain
\begin{equation}
Z_{q/p}(\epsilon)
 = \int\limits_C\!{\rm d}t\ F_p(t)
\left(1 - t {\rm e}^{-\beta(\epsilon-\mu)}\right)^{-q}
\label{zq/p}
\end{equation}
Using (\ref{Frule}), after some algebra, expression (\ref{zq/p}) can be 
transformed to
\begin{equation}
Z_{q/p}(\epsilon)
 =
\frac{1}{(q-1)!}\int\limits_0^\infty\!{\rm d}t\ t^{q-1}{\rm e}^{-t} 
\left[1+\frac{t{\rm e}^{-\beta(\epsilon-\mu)}}{p}\right]^p
\label{zq/p2}
\end{equation}

Taking $p\to\infty$ with $q$ fixed corresponds to the bosonic case. From 
(\ref{zq/p2}) we readily obtain
$$
\left.Z_{q/p}(\epsilon)\right|_{p\to\infty}
 = \left[1-{\rm e}^{-\beta(\epsilon-\mu)}\right]^{-q}
$$
which is obviously the bosonic partition function. 

Calculating (\ref{zq/p2}) in the thermodynamical limit ($q,p\to\infty$ with 
$g=q/p$ fixed) we have
\begin{equation}
Z_{q/p}(\epsilon)
 =
\left[
h^{-(1+g)/g} (h + gz)^{1/g} {\rm e}^{-1/h+1}
\right]^q
\label{zq/p3}
\end{equation}
where
\begin{equation}
h = h(g) =
\frac{1}{2}\left[1-(1+g)z + \sqrt{[1-(1+g)z]^2+4gz}\right]
\label{h}
\end{equation}
and
\begin{equation}
z \equiv {\rm e}^{-\beta(\epsilon-\mu)}
\label{z}
\end{equation}

The distribution function is defined by the relation:
\begin{equation}
n = \frac{\langle \hat N\rangle}{q} = \frac{\partial_\mu Z}{q\beta Z}
\label{distr1}
\end{equation}
From (\ref{distr1}) and (\ref{zq/p3}) we have
\begin{equation}
n = \frac{z}{h + gz}
\label{distr2}
\end{equation}
Putting in (\ref{zq/p3}) and (\ref{h}) $g=1$ we obtain
\begin{equation}
Z_1(\epsilon)
 =
\left[
h^{-2}(1) (h(1) + z) {\rm e}^{-1/h(1)+1}
\right]^q
\label{zq/p4}
\end{equation}
where
\begin{equation}
h(1) =
\frac{1}{2}\left[1-2z + \sqrt{1+4z^2}\right]
\label{h1}
\end{equation}

If we further consider the case of low densities ($z\ll 1$) we have from 
(\ref{zq/p4}):
$$
Z_1(\epsilon)
 \approx [1 + z]^q
$$
which obviously coincides with usual fermionic partition function.

Let us turn now our attention to the state-counting corresponding to the 
Haldane's dimension formula. In the case of one exclusion cell we have from 
eq.~(\ref{W(N)}):
\begin{equation}
W(N) =
\alpha(N)\sum\limits_{n_1,\dots,n_q=0}^\infty  \delta_{n_1+\dots+n_q,N}
 = \alpha(N)\cdot W_{\rm b}(N)
\label{w1}
\end{equation}
where
\begin{equation}
W_{\rm b}(N) = \frac{(q+N-1)!}{N!(q-1)!}
\end{equation}
is the bosonic statistical weight. For the set of probabilities (\ref{alpha2}) 
we obtain the following expression for the state-counting
\begin{equation}
W(N) = \frac{1}{p^N}\frac{p!}{(p-N)!}\cdot \frac{(q+N-1)!}{N!(q-1)!}
\label{w2}
\end{equation}
which is obviously different from Haldane-Wu state-counting~\cite{Hald,Wu}:
\begin{equation}
W_{\rm H}(N) = \frac{[q+(1-g)(N-1)]!}{N![q-g(N-1)-1]!}
\label{w3}
\end{equation}

\section{Haldane-Wu state-counting procedure}

Although the procedure of the last section does not lead to the combinatorial 
expression derived by Haldane and Wu, we may modify the probabilities, $\alpha$, 
so that this is obtained.

Comparing (\ref{w1}) and (\ref{w3}) we can easily write down a set of 
probabilities which provide the Haldane-Wu state-counting:
\begin{equation}
\alpha_{\rm H}(N)
 = \frac{(q-1)![q-g(N-1)+N-1]!}{(q+N-1)![q-g(N-1)-1]!}\ ,\qquad 
N = \sum_{i=1}^q n_i
\label{alh}
\end{equation}
The operator Id in this case takes the form
\begin{equation}
\mbox{Id}^{\rm H}
 = \int\limits_C\!{\rm d}t\ F_p^{\rm H}(t)
 \frac{1}{\pi^q}\int\limits\!\prod_{j=1}^q\left[{\rm d}\bar{z}_j{\rm d}z_j\ 
 {\rm e}^{-\bar{z}_j z_j} \right]
|\{z_j\sqrt{t}\}_{j=1}^q\rangle\langle \{\bar{z}\sqrt{t}\}_{j=1}^q|
\label{Idh}
\end{equation}
where
\begin{equation}
F_p^{\rm H}(t)
 = \frac{1}{2\pi i}\sum_{n=0}^p
   \frac{(q-1)![q-g(n-1)+n-1]!}{(q+n-1)![q-g(n-1)-1]!}t^{-n-1}
\label{FH}
\end{equation}

For the partition function we have the following expression:
\begin{eqnarray}
Z^{\rm H}_{q/p}(\epsilon)
 &=& 
\int\limits_C\!{\rm d}t\ F^{\rm H}_p(t)
\left(1 - t {\rm e}^{-\beta(\epsilon-\mu)}\right)^{-q}
\\
 &=& 
\sum\limits_{N=0}^p 
\frac{[q-g(N-1)+N-1]!}{N![q-g(N-1)-1]!} z^{N}
\label{zhq/p}
\end{eqnarray}
which is identical to the one used by Wu. From (\ref{zhq/p}) we can obtain the 
statistical distribution in the standard way~\cite{Wu}:
\begin{equation}
n\equiv \frac{N}{q} = \frac{1}{w({\rm e}^{\beta(\epsilon-\mu)})+g}
\label{distrh1}
\end{equation}
where function $w$ satisfies the following equation
\begin{equation}
w({\rm e}^{\beta(\epsilon-\mu)})^g 
\left[1 + w({\rm e}^{\beta(\epsilon-\mu)})\right]^{1-g}
 = {\rm e}^{\beta(\epsilon-\mu)}
\label{eqw}
\end{equation}

Turning our attention to an analog of Haldane's dimension formula for Haldane-Wu 
state-counting procedure we have from (\ref{d(N)}) and (\ref{alh}):
$$
d_{\rm H}(N) = \frac{\alpha_{\rm H}(N)}{\alpha_{\rm H}(N-1)}
 = q\frac{q-g(N-1)+N-1}{q+N-1}
 \prod_{j=1}^{N-1}\frac{q-g(N-1)+j-1}{q-g(N-2)+j-1}
$$
Taking the thermodynamical limit leads us to the following expression
\begin{equation}
d_{\rm H}(N) = \frac{q-gN+N}{1+N/q}
\label{dH}
\end{equation}
At sufficiently small densities $N/q\ll 1$ we have 
$$
d_{\rm H}(N) = q - gN + {\rm O}(N^2/q)
$$
We can conclude that at low densities Haldane's dimension and Haldane-Wu 
state-counting procedures are equivalent while in general they are not.

\section{Conclusion}

In this paper we have demonstrated the construction of coherent states for 
particles obeying Haldane exclusion statistics. This construction allows 
considerable freedom in the definition and permits the construction of states 
yielding either Haldane's dimension or Haldane-Wu state-counting. These are 
demonstrated in the limit of low densities.

These results will be used in a future publication to analyse ``non-ideal'' gas 
of particles obeying exclusion statistics.

\section*{Acknowledgments.}

This work was supported by EPSRC grant GR K68356.

\end{document}